\documentclass[iop]{emulateapj}
\usepackage{epsfig}
\usepackage{natbib}
\usepackage{amssymb}
\usepackage{amsbsy}
\usepackage{natbib}
\usepackage{subfigure}
\usepackage[mathcal]{euscript}
\usepackage{float}
\usepackage{amsmath}
\usepackage{tabularx}
\usepackage{xcolor}

\bibpunct{(}{)}{;}{a}{}{,}

\begin{document}
\title{Constraining Metallicity-dependent Mixing and Extra Mixing using [C/N] in Alpha-Rich Field Giants}
\author{
Matthew Shetrone\altaffilmark{1},
Jamie Tayar\altaffilmark{2,13,14}, 
Jennifer A.~Johnson\altaffilmark{2,3},
Garrett Somers\altaffilmark{4},
Marc H.~Pinsonneault\altaffilmark{2,3}, 
Jon A. Holtzman\altaffilmark{5}, 
Sten Hasselquist\altaffilmark{5},
Thomas Masseron\altaffilmark{6,7},
Szabolcs~M{\'e}sz{\'a}ros\altaffilmark{8,9,15}, 
Henrik J\"onsson\altaffilmark{10}
Keith Hawkins\altaffilmark{11},
Jennifer Sobeck\altaffilmark{12},
Olga Zamora\altaffilmark{6,7},
D.~A. Garc{\'{\i}}a-Hern{\'a}ndez\altaffilmark{6,7}
}
\altaffiltext{1}{University of Texas at Austin, McDonald Observatory, 32 Fowlkes Rd, McDonald Observatory, TX 79734-3005, USA}
\altaffiltext{2}{Department of Astronomy, Ohio State University, 140 W 18th Ave, Columbus, OH 43210, USA}
\altaffiltext{3}{Center for Cosmology and AstroParticle Physics, 191 West Woodruff Avenue, Ohio State University, Columbus, OH, 43210, USA}
\altaffiltext{4}{Department of Physics and Astronomy, Vanderbilt University, 6301 Stevenson Circle, Nashville, TN 37235, USA}
\altaffiltext{5}{Department of Astronomy, New Mexico State University, Las Cruces, NM 88003, USA5}
\altaffiltext{6}{Instituto de Astrof{\'{\i}}sica de Canarias (IAC),
E-38205 La Laguna, Tenerife, Spain}
\altaffiltext{7}{Departamento de Astrof{\'{\i}}sica, Universidad de La Laguna (ULL), E-38206 La Laguna, Tenerife, Spain}
\altaffiltext{8}{ELTE E\"otv\"os Lor\'and University, Gothard Astrophysical Observatory, Szombathely, Hungary}
\altaffiltext{9}{Premium Postdoctoral Fellow of the Hungarian Academy of Sciences} 
\altaffiltext{10}{Lund Observatory, Department of Astronomy and Theoretical Physics, Lund University, Box 43, SE-22100 Lund, Sweden}
\altaffiltext{11}{Department of Astronomy, the University of Texas at Austin, 2515 Speedway Boulevard, Austin, TX 78712, USA} 
\altaffiltext{12}{Department of Astronomy, University of Washington, Box 351580, Seattle, WA 98195, USA}
\altaffiltext{13}{Institute for Astronomy, University of Hawaii, 2680 Woodlawn Drive, Honolulu, Hawaii 96822, USA}
\altaffiltext{14}{Hubble Fellow}
\altaffiltext{15}{Premium Postdoctoral Fellow of the Hungarian Academy of Sciences}

\begin{abstract}
Internal mixing on the giant branch is an important process which affects the evolution of stars and the chemical evolution of the galaxy. While several 
mechanisms have been proposed to explain this mixing, better empirical constraints are necessary. Here, we use [C/N] abundances in { 26097} evolved stars 
from the SDSS-IV/APOGEE-2 Data Release 14 to trace mixing and extra mixing in old field giants with $-1.7<$ [Fe/H] $<$ 0.1. We show that the APOGEE [C/N] 
ratios before any dredge-up occurs are metallicity dependent, but that the change in [C/N] at the first dredge-up is metallicity independent for stars above [Fe/H] $\sim$ -1. We identify the position of the red giant branch (RGB) bump as a function of metallicity, note that a metallicity-dependent extra mixing episode takes place for low-metallicity stars ([Fe/H]$<-0.4$) 0.14 dex in log {\it g} above the bump, and confirm that this extra mixing is stronger at low metallicity{, reaching $\Delta [C/N] = 0.58$ dex at [Fe/H] $= -1.4$.}  We show evidence for further extra mixing on the upper giant branch, well above the bump, { among the stars with [Fe/H] $< -1.0$.  This upper giant branch mixing is stronger in the more metal-poor stars reaching 0.38 dex in [C/N] for each 1.0 dex in log {\it g}.}  The APOGEE [C/N] ratios for red clump (RC) stars are significantly higher than for stars at the tip of the RGB, suggesting additional mixing processes occur during the helium flash or that unknown abundance zero points for C and N may exist among the RC sample. Finally, because
of extra mixing, we note that current empirical calibrations between [C/N] ratios and ages cannot be naively extrapolated for use in low-metallicity stars { specifically for those above the bump in the luminosity function.}

\end{abstract}

\section{Introduction}
Carbon and nitrogen abundances on the surfaces of low-mass stars offer unique probes of their interiors. Hydrogen burning via the CNO bi-cycle occurs near enough to the surface of the star that the convective envelope brings burned material to the surface on the lower giant branch. The dredged-up material affects the star's surface C/N ratio, with the magnitude of the effect depending on the mass of the star and on any interior chemical mixing present. The former effect has been exploited to derive ages for field red giants \citep{Martig2016,Ness2016}, while the latter has been used to study so-called ``extra mixing'' in metal-poor stars \citep[e.g.,][]{Carbon1982,Charbonnel2010}. The name ``extra mixing'' has been used as a placeholder for an as-yet unconfirmed source of mixing in stars that is not included in standard stellar models. 

In standard stellar evolution theory, the initial surface abundances of carbon and nitrogen are set by the composition of the birth cloud. While diffusion can affect their surface abundances on the main sequence \citep{Richard1996}, the deepening of the convective envelope as the star expands into a red giant efficiently brings diffused elements to the surface and homogenizes the composition of the envelope at the initial abundance. As the envelope reaches depths where CN cycling has occurred, it mixes these elements into the surface convection zone, increasing the surface nitrogen abundance of these stars. This process is referred to as the first dredge-up and occurs around a surface gravity of log {\it g}$\sim 3.5$ dex. The [C/N]\footnote{We use the standard notation: [X/Q]$=logN(X)_*-logN(X)_{\odot} - logN(Q)_*+logN(Q)_{\odot}$} ratio of stars is then predicted to be constant between the end of the first dredge-up and the tip of the giant branch as well as during the subsequent core-helium-burning phase. The [C/N] ratio after first dredge-up depends on the mass of the star \citep{iben1964}, both because the depth reached by the convective envelope is larger for higher masses and because these higher-mass stars have hotter cores and thus a larger equilibrium nitrogen abundance.

These simple predictions are in conflict with observations. Analysis of star clusters \citep[e.g.,][]{Carbon1982,Kraft1994} show that additional mixing is required to match the carbon and nitrogen abundances of stars below log {\it g} of $\sim$ 2.5. This process 
is strong in metal-poor stars, but minimal in solar metallicity clusters \citep{Brown1987}. This "non-canonical extra mixing" is present in stars regardless of environment, 
and has been seen in dwarf spheroidal galaxies { \citep{shetrone2013, kirby2015, lardo2016}} and in low-metallicity field stars \citep{lambert1977}. In an influential paper, \citet[hereafter G00]{Gratton2000} used a sample of 62 low-metallicity field stars to demonstrate that the [Li/H], [C/H], [N/H], and $^{12}$C/$^{13}$C values changed not only at first dredge-up, but also at the luminosity of the "red bump," when the outwardly moving hydrogen-burning shell reaches the deepest extent of first-dredge-up. The material that participated in first dredge-up has been homogenized throughout the envelope and is therefore more hydrogen-rich than the gas previously encountered by the hydrogen-burning shell. This results in a build-up of red giants at the relevant luminosity and a bump in the luminosity function for the first-ascent red giant branch (RGB) stars. G00 also showed the core-helium-burning ``red clump'' (RC) stars had C, N, Li, and $^{12}$C/$^{13}$C values that matched the values at the tip of the RGB. In addition, O and Na did not show any trend with log {\it g}, which limits the depth of the extra mixing to above the region of the star where $O-N$ cycling occurs. 

A promising theoretical explanation was proposed by \citet{Charbonnel2010} who identified thermohaline mixing as the likely culprit. This mixing occurs because of the "salt-finger instability," when the Ledoux criteria for stability are satisfied, but a decrease in mean molecular weight ($\mu$) with depth causes the sinking of higher density material until it dissolves. In the case of red giant stars, the burning of $^3$He, via $^3$He($^3$He, 2p)$^4$He at the outer edge of the hydrogen-burning shell, reduces $\mu$. In the region where the convective envelope has never penetrated, the strong $\mu$ gradient already present from nuclear burning means that thermohaline mixing cannot begin. However, when the shell reaches the parts of the star where the penetration of the convective envelope has erased any $\mu$ gradient present, thermohaline mixing can start. Because the CNO cycle powers the hydrogen-burning shell, the shell must be hotter in metal-poor stars to provide the necessary support. Therefore the inversion of the $\mu$ gradient is deeper and hotter in metal-poor stars and the changes in C, N, and the C isotopes are much larger than for metal-rich stars. The excellent agreement between the start of extra mixing and its dependence on metallicity and the properties of thermohaline mixing means that it almost certainly plays an important role in extra mixing. 

A major criticism of the idea that thermohaline mixing is solely responsible for extra mixing is the efficiency of the mixing may need to be higher than conceivably possible. \citet{denissenkov2010} argued, based on 2D simulations, that the efficiency of mixing could be more than 10$\times$ smaller than that used by \citet{Charbonnel2010} to reproduce the observations. 3D simulations are far more appropriate \citep{garaud2015}, but have been been hampered by their computational expense and the inability to reach the low Prandtl numbers found in stars. Whether 3D simulations show efficient mixing is disputed in the literature \citep[e.g.,][]{denissenkov2011, radko2012, brown2013}. \citet{denissenkov2009} suggested that magnetic buoyancy could be combined with the thermohaline instability to create sufficient mixing, building on the work of \citet{busso2007}. This effect would also start at the red giant bump.

Additional indirect arguments raised by \citet{denissenkov2009} against thermohaline mixing as the source of extra-mixing are that it appears to be inefficient in stars that have carbon-rich material transfered to them from an asymptotic giant branch (AGB) companion and that, contrary to observations, slowly rotating red giants would be more likely to show enhanced lithium. More stringent tests of thermohaline mixing could be performed if the metallicity dependence and mass dependence could be carefully compared with models. 

However, previous results have had wide metallicity bins, small samples, or contamination by other effects present in globular clusters. The stars that G00 studied were limited in number, and were analyzed as a single metallicity bin from $-2 \leq$[Fe/H]$\leq -1$. While this sample could be expected to have similar mass, the small number of stars made it difficult to understand the trends with metallicity. The other option for samples of stars of very similar mass is RGB stars in clusters, and they have been very important in studies of the extra-mixing phenomenon.
However, globular cluster stellar populations also have light-element anomalies, such as two populations in CN band strength \citep{suntzeff81,Carbon1982} and an anti-correlation between O and Na in the abundances in individual stars \citep{sk_m5,drake1992}. 
While these anomalies are present at the main-sequence turnoff \citep{gratton2001}, well before first-dredge-up and are therefore primordial rather than evolutionary in nature, they complicate the interpretation of extra mixing in globular cluster stars.
Open clusters do not share these anomalies, but are also not present at [Fe/H]$\leq -1$ in the Galaxy and do not have many red giants in an individual cluster. To make progress interpreting extra mixing, we would therefore desire a sample 
that has good carbon and nitrogen measurements, known masses, a wide range in metallicity, and made up of field stars free from the multiple abundance populations in globular clusters. These requirements are simultaneously met by the stars in the APOGEE-2 \citep{Majewski2017,Wilson2019} survey. This analysis provides carbon and nitrogen abundances for $>$ 100,000 field red giants throughout the Galaxy. 

\citet{masseron2017} found evidence for extra mixing along the giant branch by the increased [N/Fe] ratio above the red giant bump in thin and thick disk stars throughout the entire Data Release 12 \citep{dr12} APOGEE sample. However, 
their analysis did not extensively explore the metallicity dependence of extra mixing. In addition, the thin disk covers a wide range of ages and masses, 
making the interpretation of their [C/N] evolution more complex \citep{masseron2017}. Around 2000 of the APOGEE DR12 red giants, most close to solar 
metallicity, have published masses \citep{Pinsonneault2014} and evolutionary states \citep{elsworth2017}, as a result of asteroseismic 
measurements. \citet{masseron2017} used this sample to show that higher-mass RGB stars have lower [C/N] values after first dredge-up than stars with lower 
mass, as expected. What was not expected was that the [N/Fe] values for the core-helium-burning RC stars would be $\sim$0.2 below the [N/Fe] values of the 
average on the RGB. Because the red clump stars are the immediate descendants of stars at the tip of the red giant branch, they should have similar [N/Fe] values, as indeed found by G00 for more metal-poor stars. \citet{masseron2017} proposed the dredge-up of high carbon/low nitrogen material from the helium-burning shell near the red giant tip as a possible explanation. Unfortunately, the \citet{Pinsonneault2014} sample extends only down to $\sim -0.65$ and contains relatively few stars above the RGB bump. 

For this analysis, we therefore use the Data Release 14 \citep{DR14} APOGEE-2 spectroscopic sample of red giants, which has $\sim$100,000 more stars than DR12, to map the metallicity and gravity dependence of [C/N] depletion of the RGB. We choose to analyze only stars with high [$\alpha$/Fe] values to restrict the sample in mass, guided by our understanding of stellar populations from the APOKASC sample. We divide our sample into eight metallicity bins to examine the effect of extra mixing as a function of metallicity, updating G00 for the age of large spectroscopic surveys. 

\section{Observational Data}
 APOGEE-2 is a part of the Sloan Digital Sky Survey IV \citep{Blanton2017} and is collecting high-resolution (R=22,500) spectra of $\sim$300,000 stars across the Milky Way \citep{Zasowski2017} using the Sloan Foundation Telescope \citep{Gunn2006} at Apache Point Observatory. The spectra are first run through the data reduction pipeline \citep{Nidever2015}, which includes flat-field correction, extraction of the 1D spectra, wavelength calibration, sky subtraction, and combination of spectra taken at 
different dither positions and on different nights. The resulting 1D spectra are then compared to a grid of synthetic spectra \citep{Zamora2015} using the ASPCAP analysis pipeline \citep{GarciaPerez2016} to determine the stellar parameters

\subsection{Stellar parameter and abundance measurements}
We summarize here the analysis for the parameters and elements of relevance for this study. For a more complete discussion please see \cite{Holtzman2015} and \cite{Holtzman2018}. In DR14 the ASPCAP pipeline determines the best fit, as measured by the $\chi^2$ minimum, between the entire observed spectrum and a grid of synthetic spectra. For stars with initial classification as giants, the parameters of the grid are varied in seven dimensions: T$_{\rm eff}$, log {\it g}, [M/H], microturbulence, [$\alpha$/M], [C/M], and [N/M].  The grid is fit over the entire spectra, with weights assigned based on uncertainties in individual flux values.   { The [M/H] dimension refers to the overall abundance of the model atmosphere.   The [C/M], [N/M] and [$\alpha$/M] dimensions allow for a deviation of individual classes of elements away from a scaled solar ratio for the elements C, N and alpha elements.  The alpha elements include O, Ne, Mg, Si, S, Ar, Ca, and Ti.}

The raw APOGEE surface gravities for red giants observed in the {\it Kepler} field show a systematic offset with log {\it g} determined from asteroseismology, which are accurate to 0.02 dex \citep{Hekker2013}. Using evolutionary states derived from the appearance of mixed modes in the frequency power spectrum \citep{Bedding2011}, we found that the magnitude of the offset depends on the evolutionary state. To correct the raw APOGEE gravities, we first used the seismic classifications to develop a 
robust method to  separate RC stars from red giant stars over the range in surface gravities with both evolutionary phases using spectroscopic measurements alone. We identified the T$_{\rm eff}$, as a function of log {\it g} and [M/H], that divided the RGB and the RC. For those stars within 100 K of this T$_{\rm eff}$, the [C/N] ratios were also employed to separate out the RC from the RGB \citep{Holtzman2015}.  \cite{Holtzman2018} estimate that this method has a 95\% success rate for identifying the correct evolutionary state. The corrections to the surface gravity are then applied separately to the RGB and RC samples as a function of raw 
log {\it g} and [M/H]. In this paper, when we reference surface gravity, we will always mean this corrected APOGEE log {\it g}.  

To derive abundances based on lines with negligible blending, elemental abundances are produced by fixing all but one of the stellar parameters and then comparing the spectra in small windows which are sensitive to the elemental abundance of interest. For example, for the [Fe/H] measurement we mask all but the few parts of the spectrum that are most sensitive to \ion{Fe}{1} and search for a $\chi^2$ minimum only in the [M/H] axis of the synthetic spectral grid. In a similar way, the [C/Fe] abundance 
is determined by using CO and CN windows and searching in the [C/M] axis, and the [N/Fe] abundance is determined by using CN windows and searching in the [N/M] axis. Internal systematics are removed from [Fe/H] by assuming the [Fe/H] values should be constant at all T$_{\rm eff}$ in stars in a single cluster. Because we have clusters over a range of metallicity, this gives us a correction to [Fe/H] as a function of T$_{\rm eff}$ and metallicity. The internal corrections for the raw [Fe/H] values 
are less than 0.02 dex. In this paper, ``[Fe/H]'', ``[C/Fe]'', and ``[N/Fe]'' will always refer to this corrected DR14 abundance, not to the derived stellar parameters, [M/H], [C/M] and [N/M].  { Through the rest of this work the term metallicity will refer to the corrected [Fe/H].}  It is important to note that the corrected and uncorrected abundances of [C/Fe] and [N/Fe] are identical, i.e., have not had any internal corrections applied, because the abundance ratios of these elements change within 
a single cluster for stars at different evolutionary stages on the giant branch and RC.   In this paper the derived quantity ``[C/N]'' will always refer to the difference in the elemental abundances $[C/Fe] - [N/Fe]$.

The elemental uncertainties adopted in this paper are those presented in DR14 for carbon and nitrogen. A more complete discussion of the uncertainties can be found in \citep{Holtzman2015,Holtzman2018}. A short summary is provided here for context. The elemental abundance scatter is determined from the APOGEE calibration cluster sample for individual visits and compared to the elemental abundances determined from the combined visits. This scattered is then fit by an expression as a function of 
effective temperature, metallicity, and signal-to-noise.    Each element has its own expression for the uncertainty. Thus, the uncertainty we have adopted is a good measurement of precision but does not reflect any possible systematics in accuracy as a function of effective temperature, surface gravity, or metallicity.   

\subsection{Sample Selection}
Because the [C/N] ratio on the giant branch is correlated with stellar mass, we require a sample of stars with similar mass to isolate the effects of mixing. Therefore, we chose only stars that are $\alpha$-enhanced ([$\alpha$/M]$>$ 0.14, see Figure 1). Galactic chemical evolution models predict that such stars formed before a substantial number of Type Ia supernova exploded \citep{Tinsley1979} and must therefore be old and low mass. \citet{Haywood2015}, for example, calculated isochrone-based 
ages for turnoff stars in the solar neighborhood and found that red giants with [$\alpha$/Fe] $> 0.1$ dex have ages older than $\sim$ 10 Gyr (masses less than $\sim$1.1 M$_{\odot}$). 

\begin{figure}[ht]
\centering
\includegraphics[width=.50\textwidth]{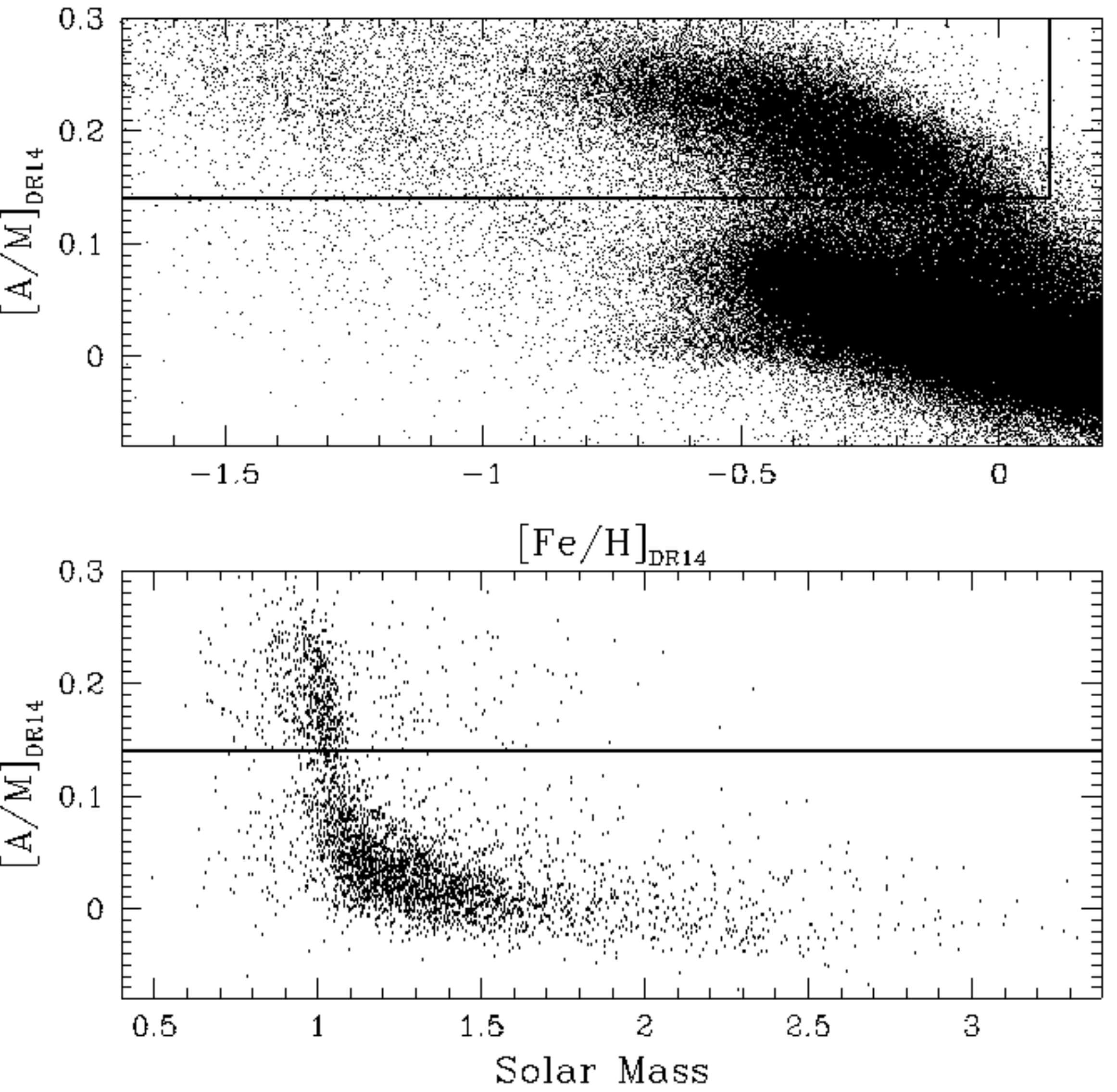}
\caption{Selection criteria for our sample and how that selection impacts the mass of our sample.  In the top panel the straight lines show the criteria that define the metal-poor ([Fe/H] $< 0.1$) $\alpha$-enhanced ([$\alpha$/M] $> 0.14$) giant stars.  In the bottom panel APOKASC sample is used to illustrate the impact of the [$\alpha$/M] selection criteria on the mass.   The second APOKASC catalog \citep{Pinsonneault2018} suggests that 12$\%$ of the alpha-rich sample may contain stars more massive than 1.2 solar masses.   }
\end{figure}

We removed all stars that were targeted as hot stars to be used for the removal of telluric absorption features (TARGFLAGS: APOGEE\_TELLURIC), all stars flagged as having "bad" or no stellar parameters (ASPCAPFLAGS: STAR\_BAD, ATMOS\_HOLE\_BAD, BAD\_PIXELS,  BAD\_RV\_COMBINATION) and all stars targeted as candidate members of globular clusters or dwarf galaxies (TARGFLAGS:  APOGEE\_SCI\_CLUSTER, APOGEE\_SGR\_DSPH, APOGEE2\_DSPH\_CANDIDATE).  To further remove potential globular cluster contaminates 
we excluded any stars if they fell within two tidal radii and had velocities within 25 km/s of the cluster mean for all clusters listed in the 2003 version of the Harris catalog \citep{Harris1996}.  The DR14 APOGEE data release does not contain any upper limits but does include errors that may be so large as to represent non-detection.  To avoid this problem, we eliminated points with very large [C/N] errors by propagating the errors in [C/Fe] and [N/Fe] and requiring the total error to be less 
than 0.3 dex.  This cut in [C/N] preferentially removes stars which are some combination of warm and metal-poor.  A total of 1085 RGB stars and 101 RC stars were rejected by this cut.  { The sample} spans a range in { [Fe/H]} from +0.1 to --2.58 and a range in surface gravity from --0.18 to 3.88. The errors on the more metal-poor stars become very significant below a { [Fe/H]} of --1.7 so we limit our analysis to above this metallicity.  { We also limit the surface gravity range to be between 3.7 
and 0.8 to avoid grid edge effects, issues with non-spherical model atmospheres, and significant AGB star contamination. Our resulting sample of alpha-enhanced stars contains 5624 RC stars and 20473 RGB stars with measured  Fe/H, alpha/Fe, log {\it g}, and [C/N]. (see Table \ref{CNdata})}  We note that the median trends in [N/Fe] are below $+0.8$ dex even among the most metal-poor stars of our final sample, well contained within the [N/M] grid, which extends from --1.0 dex to +1.0 dex.

\begin{figure*}
\centering
\includegraphics[width=.80\textwidth]{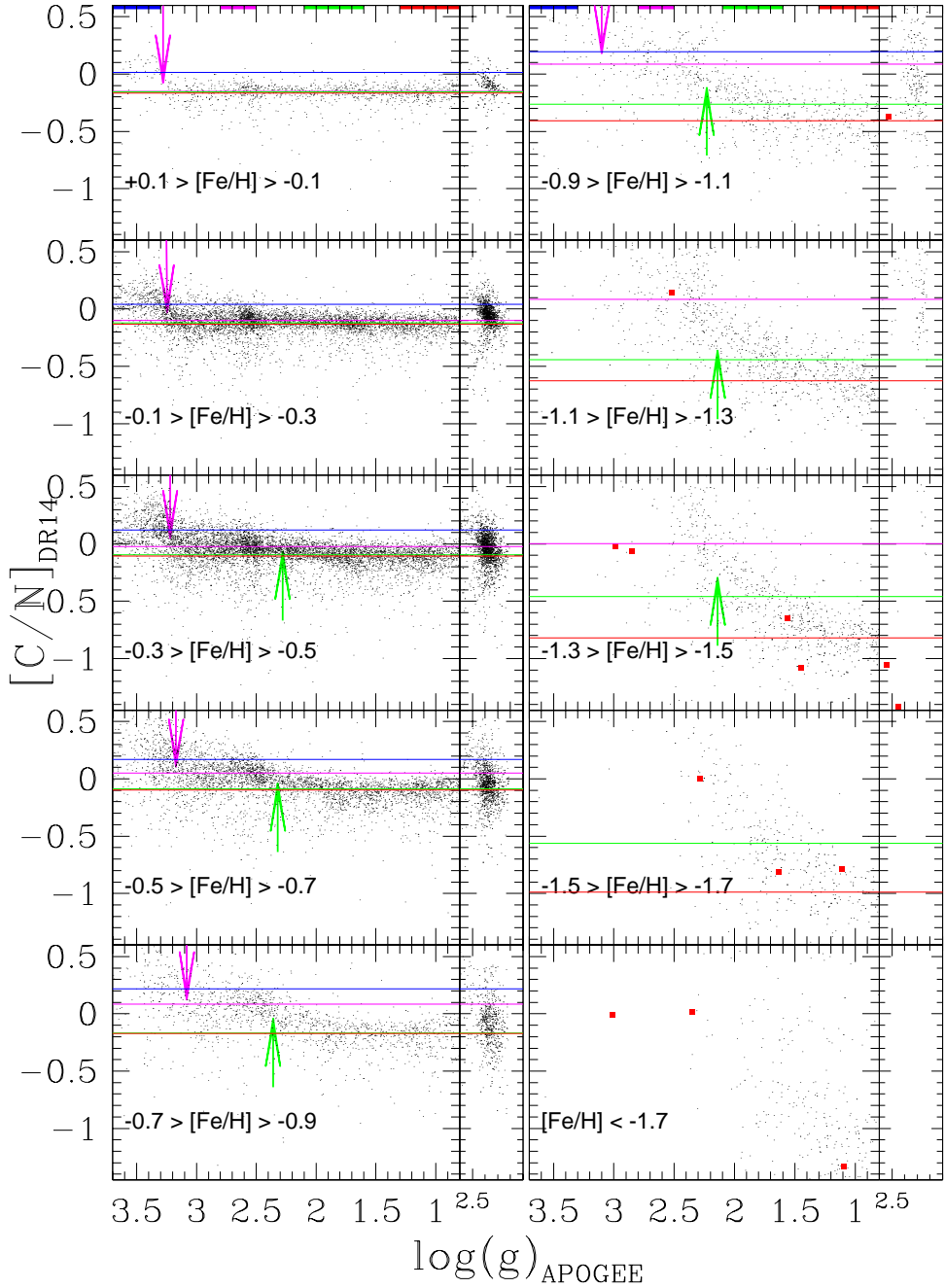}
\caption{The [C/N] ratios are shown for the RGB (left panels) and the RC (smaller right panels) samples as a function of log {\it g}. Each row is a sub-sample based on a range in { [Fe/H]} as listed on the figure.  The green and magenta arrows show the mid-point of the calculated dredge-up events as calculated and shown in Table \ref{results}.   Medians are calculated for the regions with luminosities below the first dredge-up (blue), above the first dredge-up (magenta), above the bump in the luminosity function (green), and higher up on the RGB (red). Below [Fe/H] $< -0.4$ there is some indication of extra mixing for stars with log {\it g} $<$ 2.2, consistent with this mixing occurring just above the bump in the luminosity function. Below [Fe/H] $< -0.6$ there is some indication of extra mixing for stars with log {\it g} $<$ 1.5.  { The red points represent literature [C/N] values taken from G00.} }
\label{gratton}
\end{figure*}

There are a few alpha-rich stars that are not low-mass (see, \citet[e.g.,]{Chiappini2015} for massive migrated stars, and \citet{Jofre2016}for binary mergers). However, these higher-mass alpha-enhanced stars are fairly rare.  In a DR12 APOGEE sample of RGB stars with asteroseismic masses, \citet{Martig2015} found that 6$\%$ of their alpha-rich sample had M $>$ 1.2 M$_{\odot}$. The bottom panel of Figure 1 shows stars from the second APOKASC catalog \citep{Pinsonneault2018}. This catalog combines asteroseismology from {\it Kepler} light curves with accurate temperatures of APOGEE to determine the masses. There are 6041 stars in the lower panel of which 770 have alpha abundances greater than our 0.14 dex selection criterion. The mean mass of these 770 alpha-enhanced stars is 1.033 $\pm$0.007 ($\sigma = $0.200) solar masses. There are 83 stars, 10.8$\%$ of the sample, with masses greater than 1.233 solar masses, one sigma deviation from the mean, which is greater than one might 
expect from a single-valued distribution with Gaussian errors. This is more support for the conclusions of \citet{Martig2015}. The level of contamination suggested by \citet{Martig2015} is small enough that, with some care, it should not bias any conclusions on the bulk properties of normal low-mass, alpha-enhanced giant evolution. To separate out the RGB from the RC stars we use the spectroscopic evolutionary classification discussed above. 

\section{Analysis}
  
In Figure 2, we show the evolution of the [C/N] abundance ratio in our sample stars as a function of surface gravity, similar to Figure 10 of G00. No error bars are shown in this figure because of the high density of points but the sample was chosen to have no errors larger than 0.3 dex in [C/N] and the average error among the RGB sample is 0.09 ($\sigma = 0.05$). The RC sample has been plotted separately to the right of the RGB sample. A prominent feature among the RGB sample is the high density 
of points found near log {\it g} $= 2.5$ in the more metal-rich samples. We attribute this high density of points to the bump in the luminosity function. The log {\it g} of the bump in our sample along with the mean log {\it g} of the RC are listed in Table \ref{results}. To assist in the visual interpretation of Figure 2 we have added several colored lines that represent the median [C/N] value in a small range of log {\it g} as shown at the top of the figure. The blue region was selected to be 
below the first dredge-up region in luminosity. The magenta region was selected to be above the first dredge-up but below the bump in the luminosity function. { The green line represents the median [C/N] value selected from the green region shown at the top of Figure 2 and was selected to show the [C/N] abundance after the bump and be clear of any RC contamination.  The red line represents the median [C/N] value from the red region shown at the top of Figure 2 and was selected to be the most evolved, 
luminous RGB stars in our sample.  }

 To determine the surface gravity at which the first dredge-up occurs, we fit fit a hyperbolic tangent function as a function for each metallicity bin over the range 2.6 $\leq$ log {\it g} $\geq$ 3.6 dex.    The mid-point of the hyperbolic tangent transition, the total [C/N] transition height, and the [C/N] ratio before the drop are reported in Table \ref{results}.
For the lowest metallicity bins in our sample, we lack the high-gravity stars needed to accurately measure the location of the first dredge-up and so do not report it. The lower metallicity stars, [Fe/H] $< -0.3$, exhibit an additional drop just above the luminosity bump, usually referred to as extra mixing. We fit this extra mixing with a second hyperbolic tangent 
function using stars with surface gravities between 1.5 and 2.9 dex and report the results that show a significant change in Table \ref{results}. However, the most metal-poor stars, [Fe/H] $ < -0.9$, exhibit further decline in the [C/N] ratios after the bump. Rather than a single rapid drop in [C/N], the evolution is better approximated by a linear change in [C/N] with gravity as the star ascends the giant branch. We fit this change using linear regression for stars at surface gravities between 
0.8 and 1.6 and report its slope in Table \ref{results}. 

To demonstrate more clearly the log {\it g} and metallicity dependence of the mixing, we show in Figures \ref{loggtrends} and \ref{fehtrends} fits to the [C/N] ratio as a function of log {\it g} and metallicity for stars in a number of metallicity and gravity bins. { The fits are piecewise spline and power series polynomials, of varying order depending on the degree of structure. The coefficients of these fits are listed in Tables \ref{CNlogfits} and \ref{CNFefits}.}   We also show the dispersion 
about the fits and the errors on the points, on the top and bottom of each figure, for guidance. Because the ends of the fits are poorly constrained, we have removed the first and last 20 points of the fits to avoid providing a misleading impression of the substructure. The fits in Figure \ref{loggtrends} are directly related to the panels in Figure \ref{gratton}, for example the solid purple line which starts at log {\it g} = 2.0 and drops steeply can be seen to come from the middle panel on the 
right side of Figure \ref{gratton} which has very few data points above log {\it g} $> +2.5$ but does exhibit the sharp decline in [C/N] to lower log {\it g} values. Figure \ref{fehtrends} is a different cut through the same data set and shows fits to the same data but at fixed log {\it g} bins. To give the reader a feel for the quality of the fits we provide an error of the mean for bins 0.2 dex wide in both figures. In Figure \ref{loggtrends} the top red error bars are errors in the mean for the 
$-0.1 < $ [Fe/H] $ < +0.1$ sample while the bottom black error bars are errors in the mean for the $-1.7 < $ [Fe/H] $ < -1.5$ sample.  In Figure \ref{fehtrends} the error bars represent errors in the mean for the highest and lowest surface gravity fits in bins of 0.2 dex in metallicity. Generally the error in the mean is larger at high surface gravities and at lower metallicities.  

Our final sample has four stars in common with G00.   This small number of stars in common combined with a lack of errors from the G00 sample make a direct star-by-star comparison nearly useless.  However, the G00 sample does have eight main sequence stars, 11 RGB, and three RC stars in the metallicity range $-1.95 <= $ [Fe/H] $ <= -0.92${, see the red points in Figure \ref{gratton}.}    
The G00 RGB stars' CN values are consistent with our RGB star CN values.   The G00 RC stars' CN values are systematically lower than our RC CN values and are much more in-line with our upper RGB median CN values. 

\section{Discussion}

\subsection{Mapping Extra Mixing}
Using the extensive APOGEE data set of stellar parameters and [C/N] measurements, we isolated a large sample of red giants with similar masses and $-1.7 < $[Fe/H] $< 0.1$. With this sample, we are for the first time able to trace the appearance and amount of extra mixing from metal-poor to solar-metallicity {\it field} stars.
The evolution of the depth of the extra mixing with metallicity is qualitatively consistent with the results of G00. However, the precision to which we can map the metallicity and gravity dependence of the extra mixing is substantially higher than was previously possible. The change in [C/N] at the first dredge-up looks approximately constant with metallicity for stars above [Fe/H]$>-1$. In contrast, the change in [C/N] after the luminosity bump is strongly metallicity dependent and grows smoothly 
as metallicity decreases. For the highest-metallicity stars, we see a drop in the [C/N] ratio at the location of the first dredge-up but no significant evolution of the [C/N] ratio post-dredge-up.  This is consistent with predictions and observations of clusters such as M67 \citep{Brown1987}, although we should caution the reader that nearly all open clusters with which one might compare have solar $\alpha$-element ratios unlike our $\alpha$-enhanced sample.


\subsection{Galactic Chemical Evolution of Carbon and Nitrogen} \label{evolution}
\begin{figure}

\centering
\includegraphics[width=.50\textwidth]{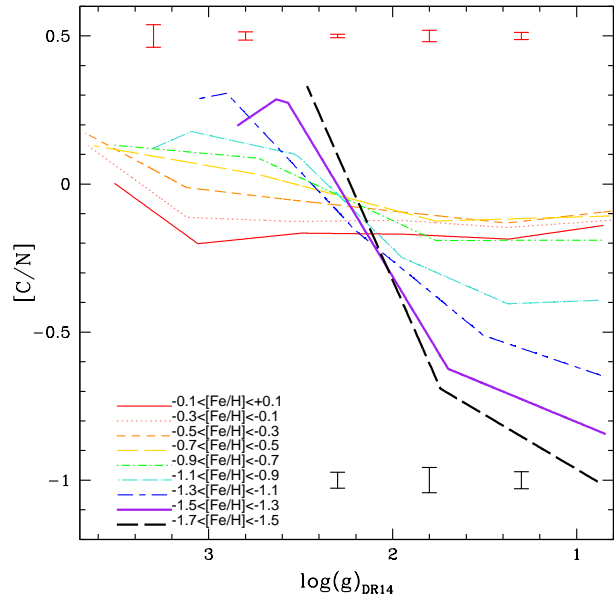}
\caption{Fits to the [C/N] ratios in metallicity ranges. Each spline fit shown in this figure represents one of the panels shown in Figure 2. The spline fit ends before the 20 lowest and highest log {\it g} points to reduce constrained excursions at the ends of the fits. The error bars along the top and bottom of the figure represent errors on the mean for bins 0.2 dex wide for the most metal-rich bin and most metal-poor bin, respectively. }
\label{loggtrends}
\end{figure}

Figure \ref{loggtrends} shows that, among the highest-gravity (least evolved) stars that have not yet undergone the first dredge-up, there is a clear trend in [C/N] with metallicity, which suggests that these elements are affected by galactic chemical evolution. This result is consistent with the results from G00 in the metallicity range $-1.1 < $[Fe/H] $< -0.9$, the most metal-rich range over which G00 has main-sequence CN ratios and the most metal-poor bin for which we have a result for the lower 
giant branch.  The average of the G00 main-sequence CN ratios in this metallicity range is super-solar, $+0.11$. This decrease in [C/N] with increasing metallicity is qualitatively similar to the predictions from \citet{VincenzoKobayashi2018} based on studies of other galaxies. However, we highlight the importance of this observation, because it implies that stellar models for modeling dredge-up and mixing should not be using a solar mixture for low-metallicity stars. This change can affect many 
of the predictions from stellar models, including the  temperature \citep{Beom2016}, lifetime \citep{Dotter2007}, and nucleosynthesis \citep{Marigo2002}.  

Another possible source for this rise would be a metallicity-dependent bias in the DR14 calculation of the $[N/Fe]$ and/or $[C/Fe]$ abundances.  { \citet{Jonsson2018} } explored biases in DR14 APOGEE abundances by comparing them against literature values {(\citet{daSilva2015}, \citet{Brewer2016}, and Gaia-ESO DR3)} on a star-by-star basis.    {  \citet{Jonsson2018} } found the carbon abundances to be in good agreement while there seemed to be a systematic trend among the nitrogen abundances.    
In top panel of Figure 6 in  { \citet{Jonsson2018}} there is an indication that either the APOGEE DR14 nitrogen abundances are { too} small as the metallicity declines or the optical abundances are too large.    However, the magnitude of the implied nitrogen bias from { \citet{Jonsson2018} } is larger than the rise in the [C/N] that we report in our analysis.   The determination of the nitrogen abundances is difficult in both the optical and the $H$-band, thus it is not clear which of these scenarios is correct. 

\subsection{The RC}
We note that the [C/N] of stars in the RC are higher than the values on the upper giant branch, suggesting that the [C/N] ratio is modified during the helium flash at the tip of the giant branch, as previously noted by \citet{masseron2017}, or that the RC stars have some hidden systematics unrecognized within the APOGEE data set. The discrepancy between the three RC stars in G00 and our results mentioned in the previous section supports the latter conclusion.   We note that the RC stars have a 
different log {\it g} correction than the RGB stars, suggesting that there is something that is not properly modeled in the ASPCAP analysis. Possible candidates for this error include errors in the carbon isotope ratio, errors in the radius for the model atmospheres themselves, or differences in the macro-turbulence for RC compared with RGB stars at a given T$_{\rm eff}$-log{\it g} combination. Some discussion of these issues can be found in \citet{masseron2017} but a full exploration of these 
effects is beyond the scope of this paper.

We also note that the [C/N] ratios of stars are used to determine the evolutionary state. This could lead to RGB stars with low [C/N] being preferentially classified as RC stars in our sample. However, [C/N] is only used in the relatively underpopulated region between the clump and the giant branch, which affects only a minor portion of our sample. Additionally, the range in [C/N] spanned by stars classified as clump and giant branch in our samples is relatively similar at fixed metallicity, suggesting it is not miss-classification driving the differences between the clump and the giant branch.


\subsection{Consequences for Ages from [C/N]}
\begin{figure}
\centering

\includegraphics[width=.50\textwidth]{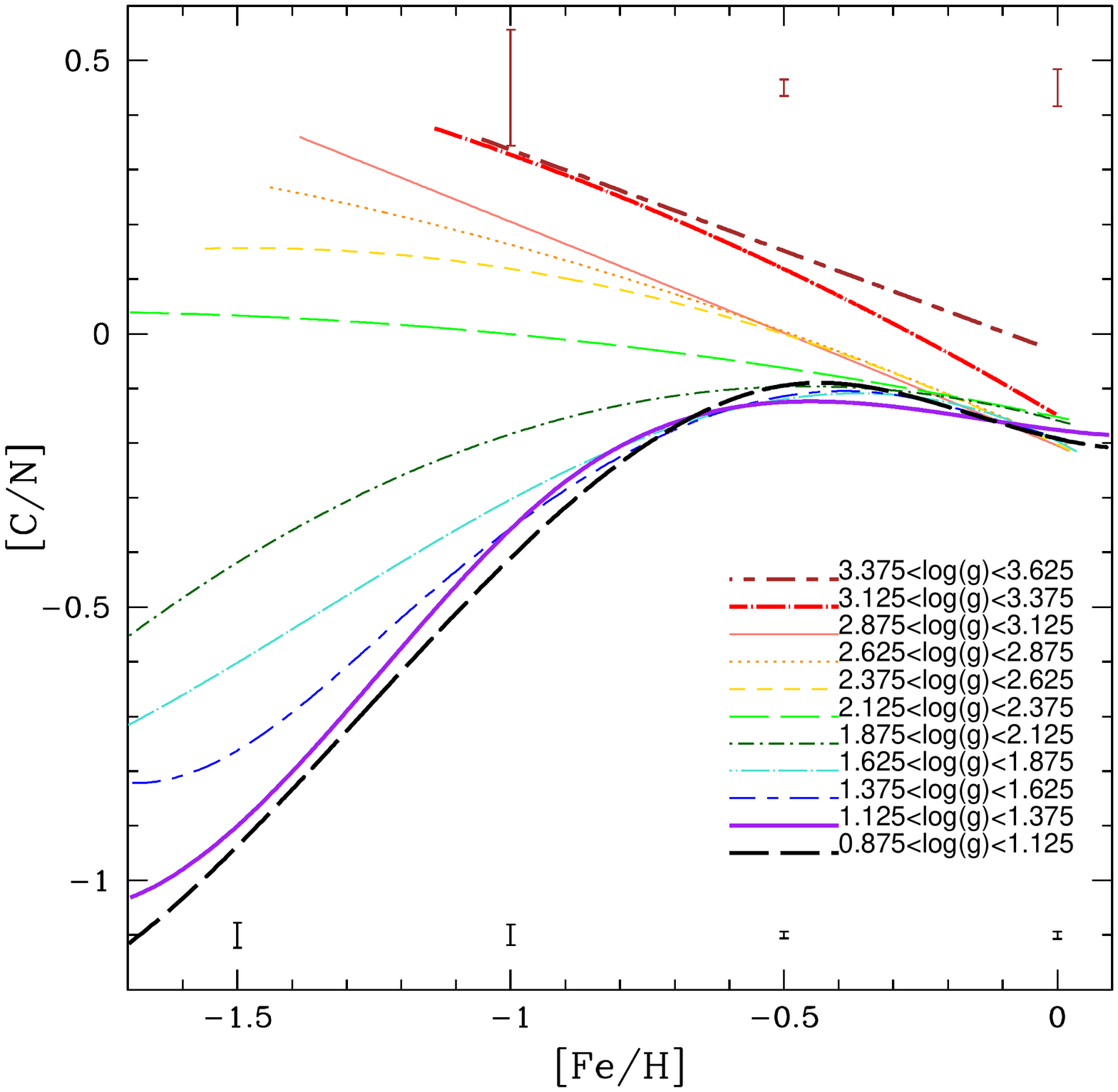}
\caption{Similar to Figure 3 except fits to the [C/N] ratios for bins of 0.25 in log {\it g} as a function of metallicity. The fit ends before the 20 lowest and highest log {\it g} points to reduce constrained excursions at the ends of the fits. The error bars along the top and bottom of the figure represent errors on the mean for bins 0.2 dex wide for the highest log {\it g} and lowest log {\it g} bins, respectively.  }
\label{fehtrends}
\end{figure}

Recent work by \citet{Martig2016} and \citet{Ness2016} has provided a calibration between the [C/N] ratio and the age of red giant stars. While these works appropriately account for the metallicity dependence of the initial [C/N] ratio and the depth of the dredge-up, the APOKASC \citep{Pinsonneault2014} sample on which they are calibrated has few stars with [Fe/H] $<$ -0.65. From Figures \ref{loggtrends} and \ref{fehtrends}, it is clear that the amount of extra mixing on the upper giant branch 
is extremely metallicity dependent, and significantly more important in stars below [Fe/H] $\sim -1$. These age calibrations therefore likely lack an explicit accounting for extra mixing, and we suggest caution when extrapolating such relationships to metal poor stars, especially those above the luminosity bump.

\begin{table*}
{
\caption{
\label{CNdata}
Sample Data\tablenotemark{a} }
\begin{tabular}{|c|c|c|c|c|c|c|c|c|}  \hline
APOGEE Name & [Fe/H] &    [A/M]       &    log {\it g}    &       [C/Fe]       &     [C/Fe]$_{err}$         &     [N/Fe]         &     [N/Fe]$_{err}$         &     classification    \\
       \hline
2M00001653+5540107 & -0.089 &  0.150 &  1.568 &  0.101 &  0.019 &  0.267 &  0.028 & RGB \\
2M00004072+5714404 & -0.248 &  0.180 &  0.966 &  0.375 &  0.021 &  0.697 &  0.030 & RGB \\
2M00004819-1939595 & -1.670 &  0.335 &  1.630 & -0.518 &  0.094 &  0.529 &  0.152 & RGB \\
2M00010088+1649201 & -0.579 &  0.231 &  2.530 &  0.043 &  0.047 &  0.067 &  0.080 & RGB \\
2M00010132+0031530 & -0.180 &  0.172 &  2.935 &  0.072 &  0.044 &  0.253 &  0.080 & RGB \\
2M00011390+6228585 & -0.447 &  0.216 &  1.674 &  0.092 &  0.027 &  0.116 &  0.041 & RGB \\
2M00011871+0011076 & -0.616 &  0.298 &  2.821 &  0.105 &  0.069 & -0.092 &  0.125 & RGB \\
2M00012134+0106579 & -0.164 &  0.185 &  3.171 &  0.054 &  0.038 &  0.165 &  0.067 & RGB \\
2M00012224+1530157 & -0.079 &  0.146 &  2.772 &  0.137 &  0.029 &  0.195 &  0.049 & RGB \\
2M00012412+6427175 & -0.260 &  0.202 &  1.023 &  0.181 &  0.017 &  0.281 &  0.024 & RGB \\
2M00012523+0012037 & -0.672 &  0.263 &  2.741 &  0.006 &  0.064 &  0.043 &  0.115 & RGB \\
2M00012984+7052497 & -0.238 &  0.190 &  2.359 &  0.157 &  0.042 &  0.129 &  0.076 & RC \\
\hline
\end{tabular}
\tablenotetext{1}{{Table 1 is published in its entirety in the machine-readable format.
      A portion is shown here for guidance regarding its form and content.}} 
      }
\end{table*}

\begin{table*}
\caption{
\label{results}
Summary [C/N] Results\tablenotemark{b} }
\begin{tabular}{|c|c|c|c|c|c|c|c|c|c|}  \hline
[Fe/H] & \multicolumn{3}{c}{First Dredge-up} & RGB Bump & \multicolumn{2}{c}{Extra Mixing} & Upper RGB Mixing   & \multicolumn{2}{c}{Red Clump} \\
       &  log{\it g} & [C/N] before &$\Delta$[C/N]             &  log {\it g}     &  log {\it g}  & $\Delta$[C/N]         & slope  [C/N]/log {\it g} & log {\it g} & $\Delta$[C/N]\tablenotemark{a}         \\
\hline
 0.0 &  3.28 $\pm$ .01&  0.02 &     0.19 & 2.54 $\pm$ .01  &  \nodata & \nodata & \nodata & 2.373 $\pm$ 0.007 &    -0.06 $\pm$ 0.01\\
-0.2 &  3.25 $\pm$ .01&  0.05 &     0.15  & 2.56 $\pm$ .01 &  \nodata & \nodata & \nodata & 2.377 $\pm$ 0.002 &    -0.08 $\pm$ 0.01\\
-0.4 &  3.22 $\pm$ .01&  0.13 &     0.15 & 2.55 $\pm$ .04 &  2.28 $\pm$ .08   &  0.06   & \nodata  & 2.386 $\pm$ 0.003 &  -0.07  $\pm$ 0.01\\
-0.6 &  3.17 $\pm$ .01&  0.17 &     0.14 & 2.52 $\pm$ .12 &  2.32  $\pm$ .11  &  0.12   & \nodata  & 2.385 $\pm$ 0.005 &  -0.06 $\pm$ 0.01\\
-0.8 &  3.08 $\pm$ .02&  0.19 &     0.14 & 2.46 $\pm$ .01 &  2.36 $\pm$ .11   &  0.21   & \nodata  & 2.371 $\pm$ 0.005 &  -0.11 $\pm$ 0.01\\
-1.0 &  3.10 $\pm$ .03&  0.27 &     0.18 & 2.30 $\pm$ .04  &  2.23 $\pm$ .09   &  0.33   & 0.08 &  2.325 $\pm$ 0.008  &   -0.39 $\pm$ 0.02\\
-1.2 &  \nodata &  \nodata &  \nodata & 2.33 $\pm$ .05 &  2.14  $\pm$ .03  &  0.52   & 0.29  &  2.290 $\pm$ 0.010   &       -0.88 $\pm$ 0.04\\
-1.4 &  \nodata &   \nodata &     \nodata & 2.25 $\pm$ .01 &  2.14  $\pm$ .03  &  0.58   & 0.25  & 2.266  $\pm$ 0.024   &    -1.28 $\pm$ 0.10 \\
-1.6 &  \nodata &   \nodata &     \nodata & 2.27 $\pm$ .07 &  \nodata & \nodata & 0.38   &  \nodata  &       \nodata  \\
\hline
\end{tabular}
\tablenotetext{1}{This compares the [C/N] of the upper RGB{, shown as the red line in Figure 2, to the median RC value.} 
\tablenotemark{b}{Errors represent the uncertainty in fitting the center of the phenomena rather that the range over which the change occurs.}}
\end{table*}

\begin{table*}
{
\caption{
\label{CNlogfits}
Piecewise Spline coefficients for the fits in Figure 3}
\begin{tabular}{|c|c|c|c|c|c|c|c|}  \hline
[Fe/H] & number of &    [C/N]$_0$       &       [C/N]$_1$       &     [C/N]$_2$         &     [C/N]$_3$         &     [C/N]$_4$         &     [C/N]$_5$     \\
       &Coefficients\tablenotemark{a}&    at log {\it g} $=0.8$   &               &                       &                       &                       &                   \\
       \hline
 0.0 &    5       &  -1.353962E-1           &     -1.860336E-1      &     -1.695335E-1      &     -1.657408E-1      &     -2.014703E-1      &      5.188016E-2 \\
-0.2 &    5       & -1.217772E-1            &     -1.468652E-1      &     -1.223153E-1      &     -1.263533E-1      &     -1.123993E-1      &      1.477976E-1 \\
-0.4 &    5       &  -9.091567E-2           &     -1.316071E-1      &     -9.388625E-2      &     -5.450294E-2      &     -1.190605E-2      &      1.817684E-1 \\
-0.6 &    3       &  -1.075440E-1           &     -1.250800E-1      &      3.257226E-2      &      1.371832E-1      &      \nodata          &     \nodata      \\
-0.8 &    3       & -1.897603E-1            &     -1.903774E-1      &      8.723510E-2      &      1.411652E-1      &      \nodata          &     \nodata      \\
-1.0 &    5       & -3.914100E-1            &     -4.053404E-1      &     -2.492756E-1      &      9.859207E-2      &      1.762891E-1      &      2.084925E-2\\
-1.2 &    4       &  -6.594786E-1           &     -5.118935E-1      &     -1.567431E-1      &      3.055076E-1      &      2.223962E-1      &     \nodata      \\
-1.4 &    3       &  -8.548605E-1           &     -6.240508E-1      &      3.019876E-1      &     -7.778014E-2      &      \nodata          &      \nodata     \\
-1.6 &    3       &  -1.036148E0            &     -6.908800E-1      &      6.320521E-1      &      2.718344E-1      &       \nodata         &     \nodata       \\
\hline
\end{tabular}
\tablenotetext{1}{{The spline extends from log {\it g} 0.8 to 3.7 dex with the first spline piece starting at log {\it g} $= 0.8$ dex.} }
}
\end{table*}

\begin{table*}
{
\caption{
\label{CNFefits}
Power Series coefficients for the fits in Figure 4}
\begin{tabular}{|c|c|c|c|c|c|c|}  \hline
 log {\it g} &   a$_{0}$    &    a$_{1}$   &       a$_{2}$   &     a$_{3}$   &       a$_{4}$   &       a$_{5}$       \\
 \hline
 1.00  &  -1.937697E-1 &  -2.100288E-1 &  9.714116E-1 &  2.788852E0   &   1.736428E0 &   3.449767E-1 \\
 1.25  &  -1.741251E-1 &  -7.799519E-2 &  4.096947E-1 & 6.844101E-1   & -1.771039E-1 &  -1.891559E-1 \\
 1.50  &  -1.942679E-1 &  -3.682857E-1 & -1.559029E-1 & 6.803854E-1  &   3.021941E-1 &  \nodata          \\
 1.75  &  -1.987884E-1 &  -5.179656E-1 & -8.219533E-1 & -1.982020E-1 &     \nodata          &  \nodata          \\
 2.00  &  -1.583757E-1 &  -2.714909E-1 & -2.965878E-1 & \nodata           &   \nodata         &  \nodata           \\
 2.25  &  -1.521781E-1 &  -2.070216E-1 & -5.558958E-2 & \nodata           &   \nodata          &  \nodata           \\
 2.50  &  -2.020961E-1 &  -4.839847E-1 & -1.631317E-1 & \nodata           &   \nodata          &  \nodata           \\
 2.75  &  -1.955374E-1 &  -4.411838E-1 & -8.280230E-2 & \nodata           &   \nodata          &  \nodata           \\
 3.00  &  -2.046856E-1 &  -4.153245E-1 & -5.783280E-3 & \nodata           &   \nodata         &  \nodata           \\
 3.25  &  -1.496861E-1 &  -5.919158E-1 & -1.153185E-1 & \nodata           &   \nodata         &  \nodata           \\
 3.50  &  -3.413932E-2 &  -3.697404E-1 & \nodata           & \nodata           &   \nodata         &  \nodata           \\
\hline
\end{tabular}
}
\end{table*}

\begin{figure*}
\centering
\includegraphics[width=.80\textwidth]{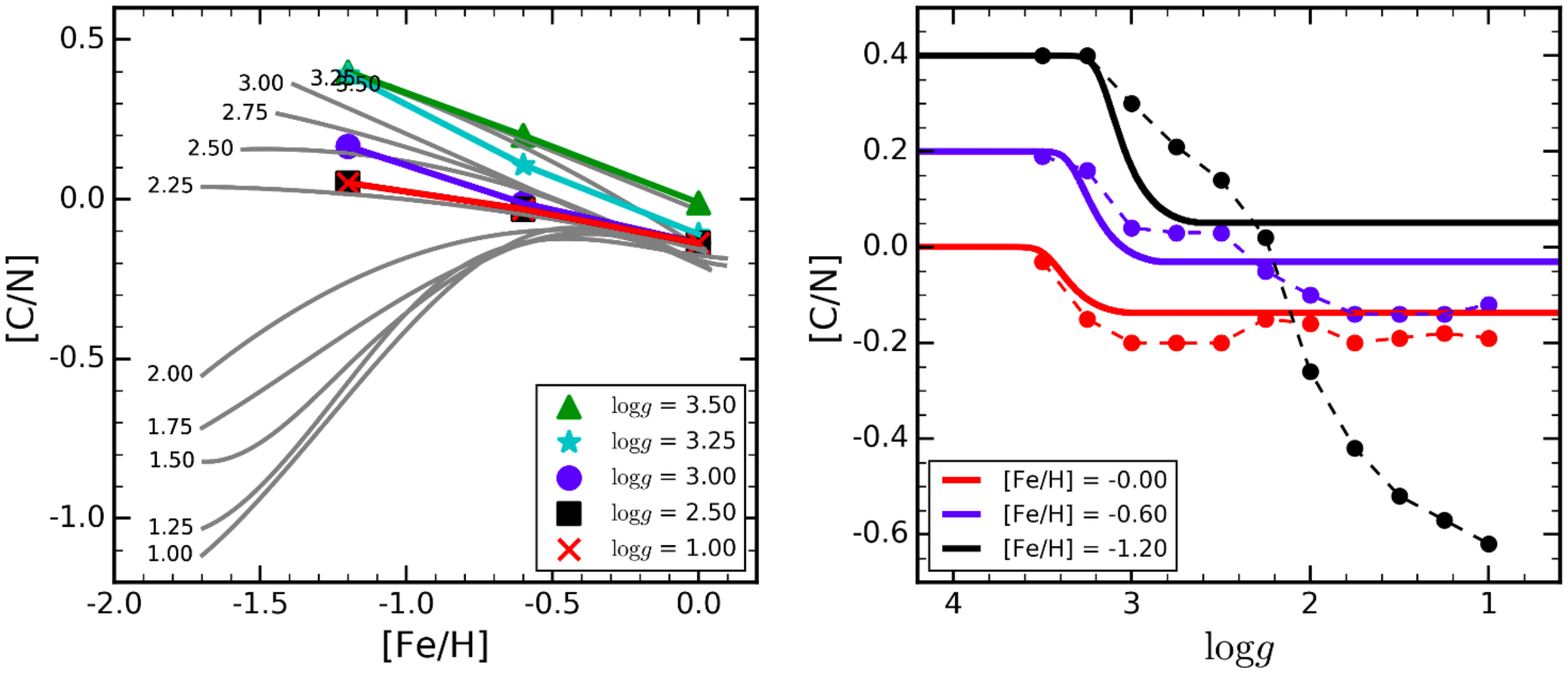}
\caption{Comparison between our [C/N] measurements and the predictions of standard models. {\it Left:} stellar models (colored lines) at several different log {\it g}s, compared to the fits from Fig. 4. The numbers in the figure label the mid-point of the bins of log {\it g} in dex from Fig 4.   The solar metallicity models predict reasonably well the dredge-up signal, but the lower-[Fe/H] models do not predict the additional mixing occurring after dredge-up. {\it Right:} surface [C/N] abundance of the models as a function of surface gravity (solid lines) compared to the fits in Fig. 3. The extra mixing signal is clearly seen in the lower-[Fe/H] models, demonstrating that the mixing signals reported in this paper are not predicted by standard stellar theory.}
\end{figure*}

\subsection{Consequences for Modeling First Dredge-up in Standard Models}

To investigate the magnitude of extra mixing relative to standard expectations, we computed 1D evolutionary models of the first ascent RGB. We use the Yale Rotating Evolution Code, adopting the fiducial physical inputs and parameters from the models of \citet{Tayar2017}. We { computed models of 0.9$M_{\odot}$, consistent with the mean value of our APOGEE sample, and at three [Fe/H] values: 0.0, -0.6, and -1.2.  These alpha-enhanced models have initial [C/N] abundances of 0.0, +0.2, and +0.4, respectively, to be consistent with the Galactic chemical evolution trends we may be seeing in our data as discussed in \ref{evolution}.

We compare these models to our data in Fig. 5. In the left panel, the gray lines show the [Fe/H] vs. [C/N] fits from Fig. 4 -- the labels indicate the mid-range log {\it g} values of each fitted sample -- and the colored lines show the results of our evolutionary models at several different log {\it g} intervals.  In the right panel, the solid lines are the results of the model predictions for the surface [C/N] abundance ratios for the three [Fe/H] values.    The dashed line and dots represent the APOGEE [C/N] values in 0.2 dex bins around these same metallicities, comparable to the fitted data in Figure 3.

\subsubsection{Agreement with Standard Models}
At solar metallicity, the models predict a dredge-up $\Delta$[C/N] of $\sim -0.1$~dex beginning around $\log {\it g} = 3.6$ and no additional depletion below log {\it g} = 3.0, in reasonable agreement with the APOGEE data.   At [Fe/H] = -0.6, the modeled first dredge-up begins at a slightly lower log {\it g} and results in a drop in the [C/N] ratio of $\sim -0.2$~dex.  The APOGEE data at this metallicity show a drop of approximately the same magnitude at the predicted log {\it g}, although further drops in the [C/N] ratio are seen in the APOGEE data and discussed in the next section.  For the lowest [Fe/H] model the onset of the first dredge-up, at log {\it g} $\sim 3.2$, is consistent with the APOGEE data but there are beyond this single point of agreement the model is a poor match for the data.

\subsubsection{Deviations from Standard Models}

Beyond the onset of first dredge-up, the standard models increasingly deviate from the observations, over-predicting the ultimate [C/N] value by 0.1~dex at [Fe/H]$ = -0.6$, and by $\sim 0.7$~dex at [Fe/H]$ = -1.2$. In the right panel of Figure 5 it is notable that the two metal-poor models make reasonable predictions until log {\it g} $\sim$~2.25, around the RGB bump, but show no additional depletion thereafter while the  APOGEE [C/N] ratio in the metal-poor stars continues to decrease.  

In summary, the standard models appear to predict the [Fe/H]-dependent log {\it g} value when dredge-up begins and roughly predicts the magnitude of the [C/N] drop down to log {\it g} $\sim$ 2.25.  The standard models do not anticipate any additional mixing after log {\it g} $\sim$ 2.25 seen among more metal-poor stars, [Fe/H] $< -0.4$.

\subsubsection{Beyond the Standard Model} }

As an additional exercise, we considered the impact that arbitrarily deep additional mixing in our models would have on the surface abundance. We found that nuclear burning in the central regions produces enough nitrogen that the observed mixing signals could be achieved if stars were fully mixed to a few pressure scale-heights below the hydrogen-burning shells. However, these calculations do not include additional [C/N] processing that would occur in envelope material mixed down into the hydrogen-burning shell, which could serve as a non-conservative source of additional [C/N] depletion. Finally, we suggest that the morphology of the continuous reduction in [C/N] at low metallicity is qualitatively consistent with more sophisticated models of diffusive mixing \citep[e.g.][]{Placco2014}, { but we postpone quantitative comparisons to specific prescriptions of extra mixing to future work.}

{
\subsection{Comparisons with dwarf galaxies}

While most analyses of dwarf galaxies exclude nitrogen or have very small sample sizes, the work of \citet{lardo2016} on the dwarf spheroidal galaxy Sculptor does offer us some small overlap in parameter space for comparison.   The bulk of the stars in \citet{lardo2016} are more metal-poor than -1.7 where we have very few stars and the most rich stars in Sculptor have oxygen abundances (which we take as a proxy for alpha enhancement) less than our $+0.14$ limit.   There are five stars which have $[Fe/H] > -1.7$, $[O/Fe] > 0.14$ and log {\it g} $> 0.8$ and these five stars are in excellent agreement with our mean trends for [C/N] as a function of log {\it g}.  If one assumes that the dwarf galaxy stars had the same ab initio [C/N] values, then from this agreement one could infer that the extra mixing that occurs in Milky Way alpha-enhanced stars also occurs in alpha-enhanced stars in other galaxies and that the Sculptor stars are very roughly the same age as our alpha-enhanced Milky Way sample.

Another dwarf galaxy which has stars with measured carbon and nitrogen is the Sagittarius dwarf galaxy \citep{hasselquist2017}.  Unfortunately, all of the \citet{hasselquist2017} sample has alpha abundance ratios less than our $+0.14$ limit; see their Figure 2.   
In Figure 4 of \citet{hasselquist2017} the $[C/N]$ ratios are shown as open blue triangles and have roughly the same values as the Milky Way low-alpha sample (blue points) and less than the high-alpha Milky Way sample (red points).   Since the comparison sample constructed by \citet{hasselquist2017} has very similar log {\it g} values and for the metallicity of the bulk of the Sagittarius stars sampled is above that for which we see significant declines of [C/N] on the upper RGB, and if one assumes that the dwarf galaxy stars had the same ab initio [C/N] values as the Milky Way sample, then we could conclude that the Sagittarius sample is younger than the alpha-enhanced sample we have constructed in this work.  

We caution the reader that nothing is known about the ab initio values of [C/N] in dwarf galaxies and this unmixed initial [C/N] could impact the mixed [C/N] along the RGB.  A full analysis of possible impacts of different starting [C/N] ratios is beyond the scope of this paper. 

}

{
\subsection{Implications for Later Stages of Stellar Evolution}

Our confirmation that extra mixing impacts stars with [Fe/H] $< -0.4 $ may have implications for later stages of stellar evolution as well.   When modeling a star as it evolves off the horizontal branch to become an AGB star, if the metallicity of the star is less than -0.4 dex then the extra mixing we see on the RGB should be factored into the initial abundance of those models.   In addition, for stars with metallicity below -0.4 dex the extra mixing we see on the RGB may play a roll on the AGB as well.   A future work could explore this with a sample of stars with different metallicities along the AGB.    The impact of such extra mixing may be very different for AGB stars than for RGB stars as there may be supplies of primary carbon and nitrogen (generated from helium burning) that are tapped by the extra mixing.    

If the extra mixing we see below [Fe/H] $< -0.4$ continues to play a roll in stellar evolution for AGB stars then we might also expect to see differences in the final abundance products for planetary nebulae (PNe) above and below this metallicity for the oldest systems.   Metal-poor models of AGB and post-AGB evolution including extra mixing, such as diffusive mixing, could potentially elucidate our understanding of PNe abundances, and our models of galactic chemical evolution, e.g. \citet{StanghelliniandHaywood2018}.

}

\section{Conclusions}
We use the alpha-rich first ascent red giants from the APOGEE field sample to map the evolution of the [C/N] ratio as a function of gravity and metallicity in a restricted mass range. { We find a discrepancy between the APOGEE [C/N] ratios for the RC stars in comparison to the tip of the giant branch [C/N] ratios.  A comparison with the few RC [C/N] values in G00 suggests that the APOGEE RC C and N abundances may be in error for RC stars. The few RGB [C/N] values in G00 are in good agreement with the APOGEE [C/N] ratios so the discrepancy appears to be limited to the RC stars.}  

We find that the [C/N] ratio before the first dredge-up is metallicity dependent, with more metal-poor stars having higher initial [C/N] values, possibly as the result of Galactic chemical evolution. We show that standard stellar evolution models can qualitatively match the first dredge-up of the stars in our sample as a function of metallicity if an appropriate, non-solar, mixture is used. 

We confirm the existence of extra mixing above the RGB bump in metal-poor stars{, [Fe/H] $< -0.4$ dex, with the mid-point of this extra mixing being 0.14 dex above the bump in the luminosity function.  We show that the amount of mixing is a smooth function of metallicity, with increasing mixing with decreasing metallicity up to $\Delta [C/N] = 0.58 $ dex at [Fe/H] $= -1.4$.} This extra mixing is not present in standard stellar evolution models, and so these models fail to reproduce the measured [C/N] in low-gravity, metal-poor stars. 

We find that either this extra mixing continues down to very low surface gravities at the metal poor end of our sample { or that a second source of extra mixing plays a roll in the upper giant branch in stars with [Fe/H] $< -1.0$.} We provide a tables of the measured [C/N] changes { as well as fits to our sample of [C/N] measurements} as a function of metallicity and gravity which can be compared to more sophisticated models including processes like thermohaline mixing in order to better constrain the mechanism responsible for this extra mixing. 

Finally, because of this extra mixing, we caution that calibrations between [C/N] and age should not be used in the low-gravity {(above the bump in the luminosity function)}, metal poor {([Fe/H] $< -0.4$)} regime unless extra mixing has been explicitly taken into account.



\begin{acknowledgements} 
We thank John Beacom for the idea of using hyperbolic tangent functions to fit the drops in [C/N]. The hospitality of the Center for Cosmology and Astroparticle Physics at The Ohio State University during the paper-writing workshop was very much appreciated. We acknowledge support from NSF grant AST-1211853. G.S. acknowledges the support of the Vanderbilt Office of the Provost through the Vanderbilt Initiative in Data-intensive Astrophysics (VIDA) fellowship.  Sz.M. has been supported by the Premium Postdoctoral
Research Program of the Hungarian Academy of Sciences, and by the Hungarian
NKFI Grants K-119517 of the Hungarian National Research, Development and Innovation Office. D.A.G.H., T.M. and O.Z. acknowledge support provided by the Spanish Ministry of
Economy and Competitiveness (MINECO) under grant AYA-2017-88254-P.  H.J. acknowledges support from the Crafoord Foundation and Stiftelsen Olle Engkvist Byggm\"astare.  The authors thank Ohio State's Center for Cosmology and AstroParticle Physics for hosting an APOGEE workshop where substantial progress was made on this project.   Support for this work was provided by NASA through the NASA Hubble Fellowship grant \#51424 awarded by the Space Telescope Science Institute, which is operated by the Association of Universities for Research in Astronomy, Inc., for NASA, under contract NAS5-26555. 
This research made use of Astropy,\footnote{http://www.astropy.org} a community-developed core Python package for Astronomy \citep{astropy:2013, astropy:2018}. 

Funding for the Sloan Digital Sky Survey IV has been provided by
the Alfred P. Sloan Foundation, the U.S. Department of Energy Office of
Science, and the Participating Institutions. SDSS-IV acknowledges
support and resources from the Center for High-Performance Computing at
the University of Utah. The SDSS website is www.sdss.org.

SDSS-IV is managed by the Astrophysical Research Consortium for the 
Participating Institutions of the SDSS Collaboration including the 
Brazilian Participation Group, the Carnegie Institution for Science, 
Carnegie Mellon University, the Chilean Participation Group, the French Participation Group, Harvard-Smithsonian Center for Astrophysics, 
Instituto de Astrof\'isica de Canarias, The Johns Hopkins University, 
Kavli Institute for the Physics and Mathematics of the Universe (IPMU) / 
University of Tokyo, Lawrence Berkeley National Laboratory, 
Leibniz Institut f\"ur Astrophysik Potsdam (AIP),  
Max-Planck-Institut f\"ur Astronomie (MPIA Heidelberg), 
Max-Planck-Institut f\"ur Astrophysik (MPA Garching), 
Max-Planck-Institut f\"ur Extraterrestrische Physik (MPE), 
National Astronomical Observatory of China, New Mexico State University, 
New York University, University of Notre Dame, 
Observat\'ario Nacional / MCTI, The Ohio State University, 
Pennsylvania State University, Shanghai Astronomical Observatory, 
United Kingdom Participation Group,
Universidad Nacional Aut\'onoma de M\'exico, University of Arizona, 
University of Colorado Boulder, University of Oxford, University of Portsmouth, 
University of Utah, University of Virginia, University of Washington, University of Wisconsin, 
Vanderbilt University, and Yale University.

\end{acknowledgements}

\bibliography{ms}
\bibliographystyle{apj} 

\end{document}